\documentclass[twocolumn]{aastex62}

\usepackage{amsmath}
\usepackage{multirow}
\usepackage{hyperref}
\usepackage{bm}

\shorttitle{Multivariate analysis of BATSE gamma-ray burst properties using skewed distributions}
\shortauthors{Mariusz Tarnopolski}

\begin{document}

\title{Multivariate analysis of BATSE gamma-ray burst properties using skewed distributions}

\author{Mariusz Tarnopolski}
\email{mariusz.tarnopolski@uj.edu.pl}

\affiliation{Astronomical Observatory, Jagiellonian University, Orla 171, 30--244, Krak\'ow, Poland}

\begin{abstract}

The number of classes of gamma-ray bursts (GRBs), besides the well-established short and long ones, remains a debatable issue. It was already shown, however, that when invoking skewed distributions, the $\log T_{90}$ and $\log T_{90} - \log H_{32}$ spaces are adequately modeled with mixtures of only two such components, implying two GRB types. Herein, a comprehensive multivariate analysis of several multi-dimensional parameter spaces is conducted for the BATSE sample of GRBs, with the usage of skewed distributions. It is found that the number of extracted components varies between the examined parameter sets, and ranges from 2 to 4, with higher-dimensional spaces allowing for more classes. A Monte Carlo testing implies that these additional components are likely to be artifacts owing to the finiteness of the data and be a result of examining a particular realization of the data as a random sample, resulting in spurious identifications.

\end{abstract}


\section{Introduction}
\label{sect1}

Gamma-ray bursts \citep[GRBs,][]{klebesadel} are separated into short \citep{eichler89,paczynski91,narayan92}, coming from double neutron star (NS-NS) or NS-black hole (BH) mergers \citep{nakar,tanvir13,abbott1,abbott2,goldstein,savchenko}, and long ones \citep{woosley93,paczynski98,macfadyen99}, whose progenitors are associated with supernovae Ib/c \citep{filippenko} connected with collapse of massive stars, e.g. Wolf-Rayet or blue supergiants \citep{galama,hjorth,stanek,woosley06,cano,perna}. The division into these two phenomenological classes was first observed in the duration distribution by \citet{mazets}. Nowadays, the threshold is set at $T_{90} \simeq 2\,{\rm s}$ \citep{kouve93}, where $T_{90}$ is the time during which 90\% of the GRB's fluence is accumulated \citep{fynbo,king,kann,li2}. The ratio of short to long GRBs is different for each satellite ($9-28\%$ of observed GRBs are short), and hence the threshold of $2\,{\rm s}$ is not robust \citep{bromberg,tarnopolski15a}. This is an instrumental effect related to the sensitivity of each instrument \citep{tarnopolski19a}.

Since it was noted that $\log T_{90}$ can be approximately described by a mixture of normal distributions \citep{mcbreen,koshut}, the number of GRB classes was most often inferred based on such an approach. This lead \citet{horvath98} to report on the presence of a third peak in the sample of 797 GRBs detected by the Burst And Transient Source Explorer onboard the {\it Compton Gamma-Ray Observatory} \citep[{\it CGRO}/BATSE;][]{meegan92,paciesas}. However, with more data accumulated\footnote{There are more than 2000 GRBs in the complete BATSE catalog.}, this peak nearly disappeared, adding to the asymmetry (skewness) of the long GRB group \citep{horvath02,tarnopolski15b}. Nevertheless, the existence of a third normal component in the distribution of $\log T_{90}$ was also claimed later for data gathered by the {\it Swift} Burst Alert Telescope \citep{horvath08,horvath10,zhang08,huja,zitouni,zhang16}. On the other hand, it was found that two and three Gaussian components are required in the rest and observer frames, respectively, for the {\it Swift} GRBs \citep{huja,tarnopolski16b,zhang16,kulkarni}. \citet{zitouni}, though, found that three groups are required in both frames; adidtionally, only two Gaussian components turned out to be needed to adequately describe the BATSE dataset in the observer frame \citep{zitouni,zhang16}, contrary to the findings of \citet{horvath02}. In case of {\it Fermi} Gamma-ray Burst Monitor \citep{kienlin,gruber,bhat}, it was found that two Gaussian components suffice for the $\log T_{90}$ distribution to be appropriately modeled \citep{bystrycky,bhat,zhang16,kulkarni}. The same conclusion was reached by \citet{zitouni18} by employing pseudo-redshifts.

It has been already argued though \citep{koen,tarnopolski15b,tarnopolski19a} that the observed $\log T_{90}$ distribution need not be actually comprised of normal components. Modeling an inherently skewed distribution with a mixture of symmetric ones (e.g. Gaussian) inevitably requires excessive components to be included, resulting in a spurious determination of the number of classes. The observed asymmetry can originate from, e.g., an asymmetric distribution of the progenitor envelope mass \citep{zitouni}, or from convolution with the redshift distribution \citep{tarnopolski19a}. Therefore, the BATSE, {\it Swift} and {\it Fermi} univariate $\log T_{90}$ data sets were examined previously with various skewed distributions \citep{tarnopolski16a,tarnopolski16c,kwong}, as it is conceptually simpler to introduce an additional parameter in modeling of the two well established groups of GRBs rather than to construct a new physical scenario giving rise to the elusive intermediate class. It was indeed found that mixtures of only two skewed components are either significantly better than, or at least as good as, three-component symmetric models, meaning that the third class is unnecessary \citep{tarnopolski16a}.

Higher-dimensional parameter spaces are a natural next step in dividing GRBs into physically meaningful classes, starting from the two-dimensional realm of the $T_{90}-H$ plane composed of the duration and hardness ratio. \citet{horvath06} (utilizing BATSE data), \citet{ripa09} (with {\it RHESSI}), \citet{horvath10,veres} (using {\it Swift} GRBs) performed analyses similar as in the univariate case, i.e. assumed a bivariate Gaussian mixture model and seeked the number of components fitting the data best. They all found a model with three components to be more favorable than that with two components. \citet{ripa12}, however, for the {\it RHESSI} data arrived at only two components. Likewise, \citet{yang} examined a sample of {\it Swift} GRBs with measured redshifts, and also showed that two components suffice, in both the observer and rest frames. For the {\it Fermi} sample, contradictory results have been obtained: \citet{bhat} arrived at two, while \citet{horvath18} at three components as the most favorable scenario.

Finally, a consequence of the results of \citet{tarnopolski16a,kwong} was to exploit bivariate skewed distributions in the $T_{90}-H$ plane \citep{tarnopolski19a}, and lead to concluding that the BATSE and {\it Fermi} data sets are also sufficiently well described by mixtures of only two skewed components. GRBs from {\it Swift} \citep{lien16}, {\it Konus}-Wind \citep{svinkin16}, {\it RHESSI} \citep{ripa09}, and {\it Suzaku}/WAM \citep{ohmori} yielded less unambiguous conclusions \citep{tarnopolski19b}, although no clear evidence for a mixture of three components was found.

On several occasions, parametric clustering of GRBs was done in parameter spaces with dimension greater than two. \citet{mukh} performed multinormal modeling\footnote{Following non-parametric clustering in a six-dimensional space that yielded ambiguous results, pointing at either two or three groups.} (in the space of $T_{90} - F_{\rm tot} - H_{321}$) that indicated three components in the GRB population. For the {\it RHESSI} data \citep{ripa12} multinormal fitting in the three-dimensional space of $T_{90}$, $H$ and peak-count rates also yielded three components. \citet{chattopadhyay17} examined the complete BATSE data in a six-dimensional space\footnote{\label{foot}The same as in \citep{mukh} for the non-parametric clustering. It should be noted that $T_{90}$ and $T_{50}$ are highly correlated, so are $H_{32}$ and $H_{321}$ (see Sect.~\ref{sect2.1} herein).} (composed of durations $T_{90}$ and $T_{50}$, peak flux measured in 256 ms bins $P_{256}$, total fluence $F_{\rm tot}$, and hardness ratios $H_{32}$ and $H_{321}$). By means of a multinormal mixture model, they arrived at a conclusion that there are five clusters in this space. The same result was achieved by modeling with a multivariate Student-$t$ distribution \citep{chattopadhyay18}.

\citet{acuner} employed the Gaussian mixture model to analyze {\it Fermi} GRBs in a different five-dimensional space of the \citet{band} spectral parameters $(\alpha,\beta,E_{\rm peak})$, the duration $T_{90}$ and the fluence, and also claimed evidence for five groups. For completeness, it should be noted that \citet{ruffini18} specified seven GRB classes.

In order to constrain the number of variables to a meaningful subset, \citet{bagoly98} performed a thorough principal component analysis (PCA) on 625 BATSE GRBs, employing nine variables: the durations $T_{90}$ and $T_{50}$, fluences $F_1$, $F_2$, $F_3$, $F_4$, and peak fluxes $P_{64}$, $P_{256}$, $P_{1024}$. They concluded that the two first PCs constitute over 90\% of the information. Hence, the BATSE GRBs could be well described by only two variables, constructed from the nine observables (with most weight on the duration, peak flux, and total fluence, among which only two are independent). Including also the third PC (while not necessary from the point of view of statistical theory) elevates the information content in the three PCs to nearly 97\%. The four fluences can be expressed by the total fluence and hardness ratios; then, $H_{43}$ is a significant component for the PCs. These results hold for a bigger sample of 1598 GRBs (\citealt{balastegui}; see also Sect.~\ref{sect2.1} herein). \citet{balazs03} built on this PCA analysis, and argued for astrophysically different origin of short and long GRBs by investigating the $T_{90}-F_{\rm tot}$ relation for BATSE GRBs. \citet{horvath12} performed PCA on a sample of 425 {\it Fermi} GRBs that was followed by fitting mixtures of multinormal distributions in a three-dimensional PC space, and found that a three-component model is the optimal one in terms of goodness of fit.

Most recently, \citet{toth19} performed the usual Gaussian mixture modeling of the BATSE catalog, utilizing six variables\footnote{See footnote \ref{foot}.}: $T_{90}$, $T_{50}$, $F_{\rm tot}$, $P_{256}$, $H_{32}$, and $H_{321}$. They found that this six-dimensional space is best described by five Gaussian components, consistent with \citep{chattopadhyay17}. This is due to a skewed $P_{256}$ distribution, which leads to cutting two, otherwise also skewed, clusters (corresponding to short and tentative intermediate GRBs) into two, resulting in four clusters identified by the Gaussian model, and the single group for long GRBs.

The GRB family, besides the short and long ones, includes also ultra-long GRBs \citep{gendre,levan,zhang14,perna}, low-luminosity GRBs \citep{bromberg11}, and short GRBs with extended emission \citep[sGRBwEE;][]{norris06} i.e. having durations that would classify them as long GRBs, but without an associated supernova; they most likely originate from the merger of a white dwarf with an NS \citep{king} or BH \citep{dong}. Further separation of long GRBs into subgroups was also considered \citep{pendleton97}, with sGRBwEE as one of the subgroups \citep{tsutsui13a,tsutsui14}. A contamination of the BATSE sample of short GRBs by soft gamma repeaters, on the level of up to a few per cent, was also considered \citep{lazzati05,ofek07,ofek08,hurley10}. A higher representation of sGRBwEE in the catalogues \citep{bostanci13,kaneko15,kagawa19} is more challenging to take account for. What actually constitutes a GRB class remains a matter of debate, though.


The aim of the presented work is to perform a possibly exhaustive multivariate analysis of the {\it CGRO}/BATSE data in several parameter spaces with dimensions ranging from one to four, in order to establish the number of GRB classes.  In Sect.~\ref{sect2} the examined data sets and statistical methods are briefly described. In Sect.~\ref{sect3} the reliability of the fitting is investigated using simulated datasets. Sect.~\ref{sect4} presents the results. Sect.~\ref{sect5} is devoted to discussion and gathers concluding remarks. The \textsc{R} software\footnote{\url{http://www.R-project.org/}} is utilized throughout; the fittings are performed with the package \textsc{mixsmsn}\footnote{\url{https://cran.r-project.org/web/packages/mixsmsn/index.html}} \citep{prates}, and the \textsc{Mathematica} computer algebra system is also utilized.

\section{Datasets and methods}
\label{sect2}

\subsection{Samples}
\label{sect2.1}

{\it CGRO}/BATSE current catalog\footnote{\url{https://heasarc.gsfc.nasa.gov/W3Browse/all/batsegrb.html}} contains in total 2702 GRBs, and for several of them provides the following observables\footnote{Only for a handful of BATSE GRBs the redshifts could be measured from the afterglow observations \citep{bagoly03}. Nowadays, the number of redshift measurements, among various satellites, is about 600 \citep{wang19}.}: durations $T_{90}$ and $T_{50}$, fluences $F_1 = F_{20-50\,{\rm keV}}$, $F_2 = F_{50-100\,{\rm keV}}$, $F_3 = F_{100-300\,{\rm keV}}$, $F_4 = F_{>300\,{\rm keV}}$, and peak fluxes on 64, 256, and 1024 ms time scales, $P_{64}$, $P_{256}$, $P_{1024}$. The hardness ratios utilized herein are defined as $H_{32}=F_3/F_2$, $H_{43}=F_4/F_3$, $H_{321}=F_3/(F_2+F_1)$. 

For completeness, the univariate $\log T_{90}$ distribution is also examined. An analysis using different skewed distributions was performed previously \citep{tarnopolski16a}. The $\log T_{90} - \log H_{32}$ plane of 1954 GRBs was also analyzed \citep{tarnopolski19a}, but because $\log T_{90} - \log H_{43}$ is examined herein for $F_4\neq 0$ (1598 GRBs), the same condition was applied for the sample of $\log T_{90} - \log H_{32}$ to be reanalyzed (1597 GRBs). Next, the PCA from \citep{bagoly98} was repeated for the current catalog (1598 GRBs with all 9 variables; see also \citealt{balastegui,borgonovo06}). PCA aims to make a linear combination of the variables in a data set in such a way that the variance contained in only a few first PCs accounts for most of the information conveyed in the data \citep{dunteman}. It is used to reduce the dimensionality of the system under investigation. Remarkably, almost 1000 GRBs more than analyzed by \citet{bagoly98} do not affect the the outcomes of PCA. Specifically, the elements of the correlation matrix from Table~3 of \citet{bagoly98} do not differ by more than 0.03, leading to nearly the same PCs as in Table~4 of \citet{bagoly98}. Therefore, all conclusions of \citet{bagoly98} still hold. In particular, the first two PCs constitute 91.3\% of the information content in the current catalog. The third PC adds 5\%, leading to a total information content of 96.3\%. Therefore, the spaces $PC_1-PC_2$ and $PC_1-PC_2-PC_3$ are sufficient, and are examined here as well.

From the PCA it follows that in the roughest approximation the first two PCs correspond to the duration and total fluence, hence the  $\log T_{90} - \log F_{\rm tot}$ plane is examined as well (see also \citealt{balazs03}). Two cases are considered: rejecting GRBs with $F_4 = 0$ (1598 GRBs), and including those with $F_4 = 0$ (1927 GRBs). The third PC is roughly identical to $F_4$, so a three-dimensional space of $\log T_{90} - \log F_{\rm tot} - \log F_4$ is analyzed. Because of singling out $F_4$, and the fact that $H_{43}$ appears to have larger variance than $H_{32}$, the $\log T_{90} - \log F_{\rm tot} - \log H_{43}$ space is examined as well. Finally, \citet{toth19} recenly conducted multidimensional Gaussian mixture modeling on six parameters: $T_{90}$, $T_{50}$, $F_{\rm tot}$, $P_{256}$, $H_{32}$, and $H_{321}$. However, $\log T_{90}$ and $\log T_{50}$ are highly correlated (Pearson coefficient $r=0.967$), as well as $\log H_{32}$ and $\log H_{321}$ ($r=0.961$), therefore herein a four-dimensional space of $\log T_{90} - \log F_{\rm tot} - \log H_{32} - \log P_{256}$ is taken into account (where GRBs with $F_4 = 0$ are allowed). To summarize, the following datasets are considered herein:
\begin{enumerate}
\item $\log T_{90}$ (2037 GRBs),

\item $PC_1-PC_2$ (1598 GRBs),
\item $PC_1-PC_2-PC_3$ (1598 GRBs),

\item $\log T_{90} - \log H_{43}$ (1598 GRBs),
\item $\log T_{90} - \log H_{32}$ (1597 GRBs),

\item $\log T_{90} - \log F_{\rm tot}$ (1598 GRBs),
\item $\log T_{90} - \log F_{\rm tot}$ (1927 GRBs),

\item $\log T_{90} - \log F_{\rm tot} - \log F_4$ (1598 GRBs),
\item $\log T_{90} - \log F_{\rm tot} - \log H_{43}$ (1598 GRBs),

\item $\log T_{90} - \log F_{\rm tot} - \log H_{32} - \log P_{256}$ (1927 GRBs).
\end{enumerate}

\subsection{Methodology}
\label{sect2.2}

The methodology is the same as in \citet{tarnopolski19a,tarnopolski19b}. Mixtures of the following multivariate distributions are fitted: multinormal (Gaussian, G), skew-normal (SN), Student $t$ (T), and skew-Student (ST). The multinormal distribution is characterized by the location (the mean) $\bm{\mu}$, and the covariance matrix $\bm{\Sigma}$, with elements composed of the standard deviations, $\bm{\sigma}$, and correlations, $\bm{\rho}$. A mixture of $n$ components is characterized by $p=6n-1$ parameters. The SN distribution admits an additional skewness parameter, $\bm{\lambda}$. In total it requires $p=8n-1$ parameters. The T distribution is also characterized by $\bm{\mu}$ and $\bm{\Sigma}$, and by the number of degrees of freedom, $\nu$. This is a symmetric distribution, but with higher kurtosis (leptokurtic, heavier tails) than the Gaussian. It is described by $p=6n$ parameters. The ST distribution admits the skewness parameter, $\bm{\lambda}$. It yields a total of $p=8n$ parameters. When $\bm{\lambda}=0$, the SN and ST reduce to G and T distributions, respectively. The T distribution becomes a Gaussian when $\nu\rightarrow\infty$. 

Fitting the mixture distributions is performed by maximizing the loglikelihood, $\mathcal{L}$. The fits are compared using the small sample\footnote{It is advised \citep{hurvich89,burnham} to employ the small sample correction when $N/p<40$, which is the case for a few instances herein.} Akaike and Bayesian Information Criteria ($AIC_c$ and $BIC$), given by
\begin{equation}
AIC_c=2p-2\mathcal{L}+\frac{2(p+1)(p+2)}{N-p-2}
\label{}
\end{equation}
and
\begin{equation}
BIC=p\ln N-2\mathcal{L},
\label{}
\end{equation}
with $N$ data points \citep{akaike,schwarz,hurvich89,burnham,biesiada,liddle,tarnopolski16b,tarnopolski16a,tarnopolski19a,tarnopolski19b}. A preferred model is the one that minimizes $AIC_c$ or $BIC$. $AIC_c$ is liberal, and has a tendency to overfit, i.e. it might point at an overly complicated model in order to follow the data better. $BIC$ is much more stringent, and tends to underfit, i.e. it prefers models with a smaller number of parameters. Therefore, when the two $IC$ point at different models, the truth lies somewhere in between. (See \citealt{tarnopolski19a} and references therein for more details.)

What is essential in assesing the relative goodness of a fit in the $AIC_c$ method is the difference, $\Delta_i=AIC_{c,i}-AIC_{c,\rm min}$, between the $AIC_c$ of the $i$-th model and the one with the minimal $AIC_c$. If $\Delta_i<2$, then there is substantial support for the $i$-th model (or the evidence against it is worth only a bare mention), and the proposition that it is a proper description is highly probable. If $2<\Delta_i<4$, then there is strong support for the $i$-th model. When $4<\Delta_i<7$, there is considerably less support, and models with $\Delta_i>10$ have essentially no support \citep{burnham,biesiada}.

In case of $BIC$, $\Delta_i=BIC_i-BIC_{\rm min}$, and the support for the $i$-th model (or evidence against it) also depends on the differences: if $\Delta_i<2$, then there is substantial support for the $i$-th model. When $2<\Delta_i<6$, then there is positive evidence against the $i$-th model. If $6<\Delta_i<10$, the evidence is strong, and models with $\Delta_i>10$ yield a very strong evidence against the $i$-th model \citep[essentially no support;][]{kass}.

\section{Reliability of the fits}
\label{sect3}

To establish how accurately the fitting can recognize the true underlying distribution, 100 realizations of 2000 points each were generated from mixtures of two trivariate Gaussian components, with parameters drawn from uniform distributions: $\mu_a \sim \mathcal{U}(0,1)$, $\mu_b \sim \mathcal{U}(-1,0)$, $\sigma \sim \mathcal{U}(0.1,0.9)$, $\rho \sim \mathcal{U}(-0.9,0.9)$, $p \sim \mathcal{U}(0.2,0.5)$. Only draws leading to positive definite covariance matrices $\bm{\Sigma}$ were kept for further calculations. Next, the 2G, 3G, 2SN and 3SN models were fitted for each realization and the best, according to $AIC_c$ and $BIC$, was recorded. Pairwise comparisons with the 2G fits were performed, and $AIC_c$ implied that:
\begin{itemize}
\item 3G was better than 2G in 23\% of cases,
\item 2SN in 8\%,
\item 3SN in 22\%,
\end{itemize}
while $BIC$ pointed unambiguously at 2G in 100\% of cases.

The same methodology was applied to realizations from 2SN distributions, with the shape parameters $\lambda \sim \mathcal{U}(-10^3,10^3)$ The outcome turned out to be more involved. According to $AIC_c$:
\begin{itemize}
\item 2G was better than 2SN in 7\% of cases,
\item 3G in 17\%,
\item 3SN in 30\%.
\end{itemize}
$BIC$ yielded similar results:
\begin{itemize}
\item 2G in 8\% of cases,
\item 3G in 16\%,
\item 3SN in 26\%.
\end{itemize}

Hence when non-skewed distributions are considered, $BIC$ is very accurate, while $AIC_c$, due to its tendency to overfit, might point at an incorrect model, in particular---it may indicate the presence of more components than are actually present. For skewed distributions, both $AIC_c$ and $BIC$ are consistent with each other, yet they also consistently indicate an incorrect model. Globally, among the four models examined, for the underlying 2G distribution, $AIC_c$ pointed at:
\begin{itemize}
\item 2G in 50\% of cases,
\item 3G in 27\%,
\item 2SN in 8\%,
\item 3SN in 15\%,
\end{itemize}
while $BIC$ yielded 2G in all cases. For the underlying 2SN, $AIC_c$ gave:
\begin{itemize}
\item 2G in 0\% of cases,
\item 3G in 3\%,
\item 2SN in 66\%,
\item 3SN in 31\%,
\end{itemize}
and $BIC$ indicated:
\begin{itemize}
\item 2G in 0\% of cases,
\item 3G in 3\%,
\item 2SN in 73\%,
\item 3SN in 24\%.
\end{itemize}

\section{Results}
\label{sect4}

\begin{figure}
\includegraphics[width=\columnwidth]{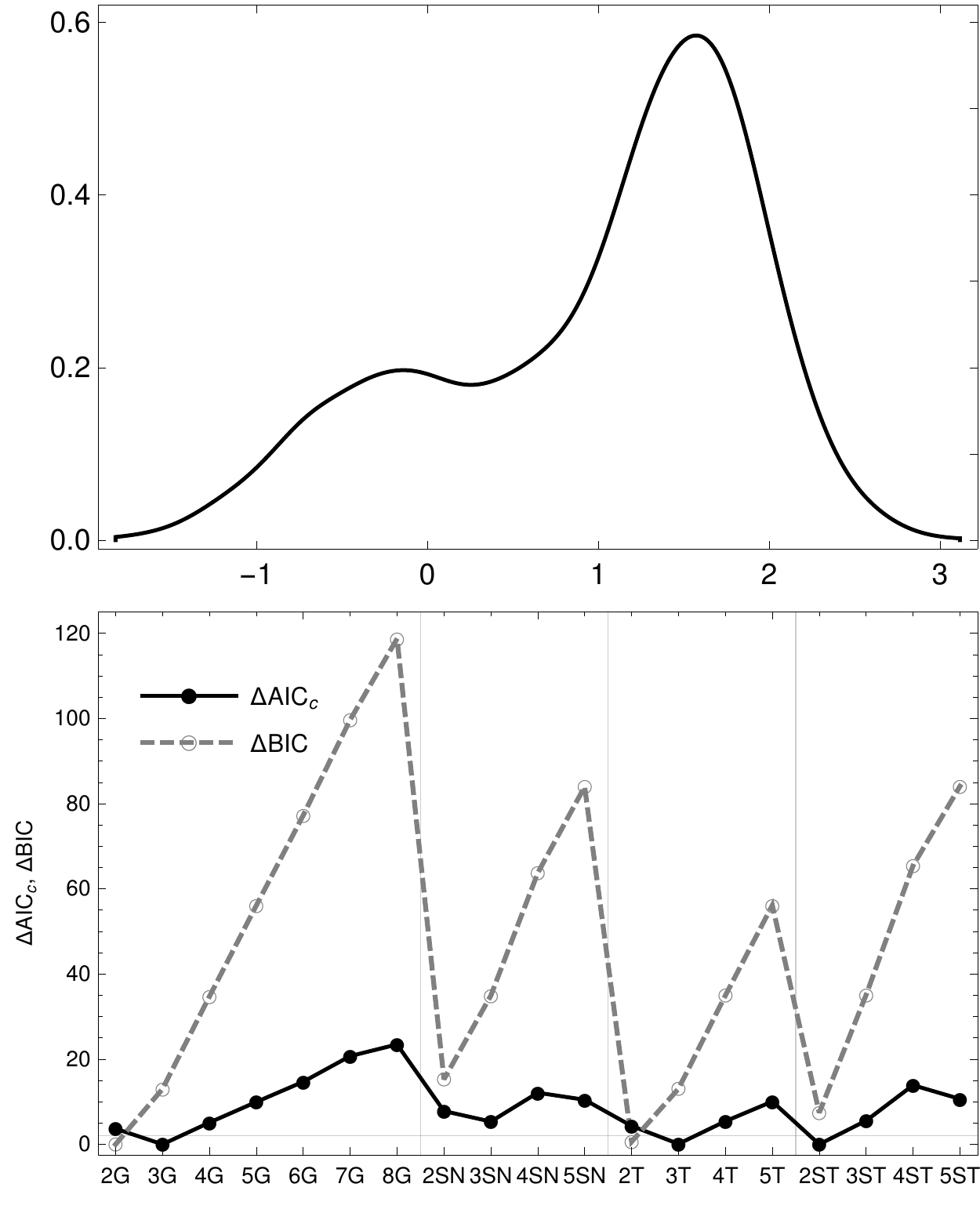}
\caption{Histogram (top) and information criteria scores (bottom) for $\log T_{90}$.}
\label{fig0}
\end{figure}

\begin{figure}
\includegraphics[width=\columnwidth]{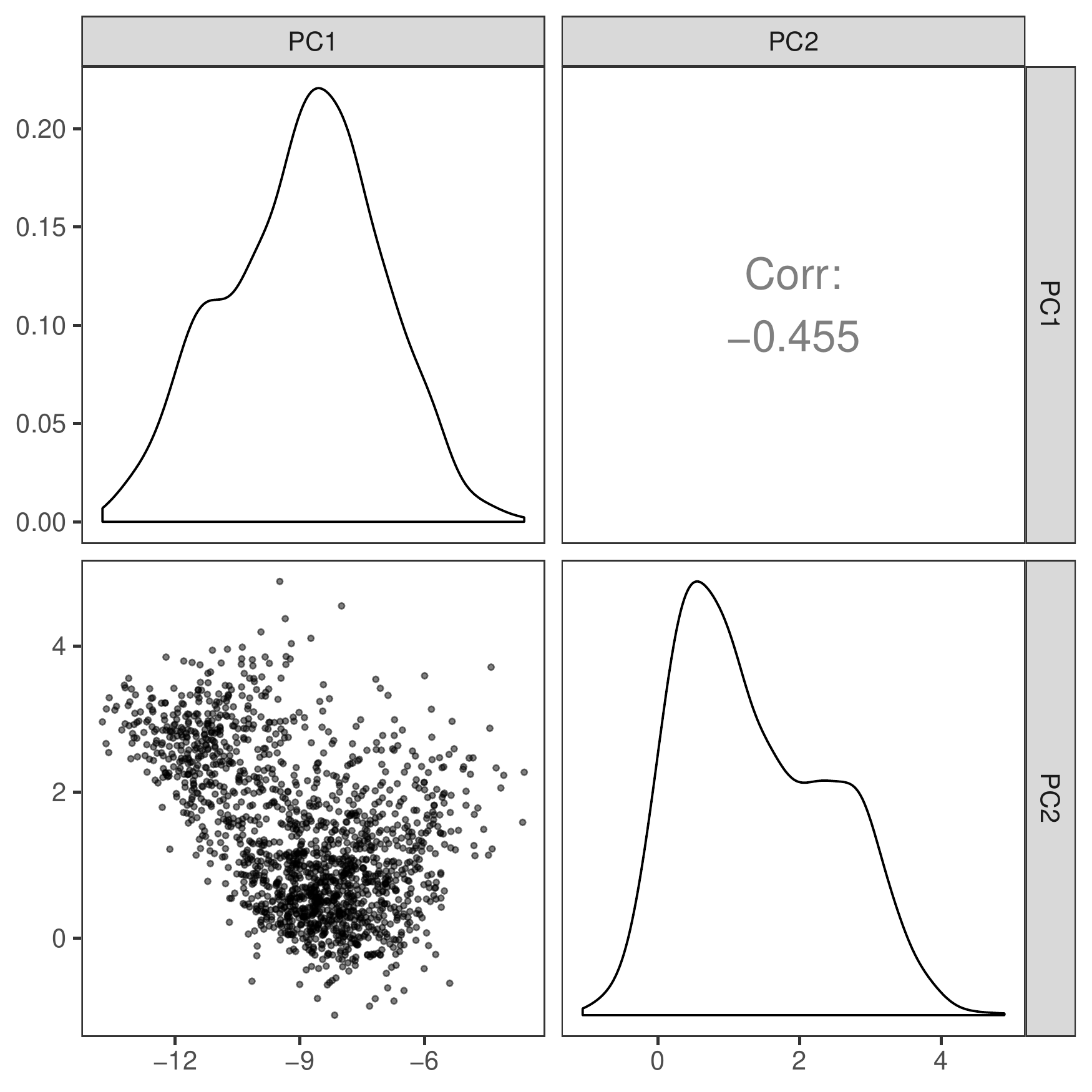}
\includegraphics[width=\columnwidth]{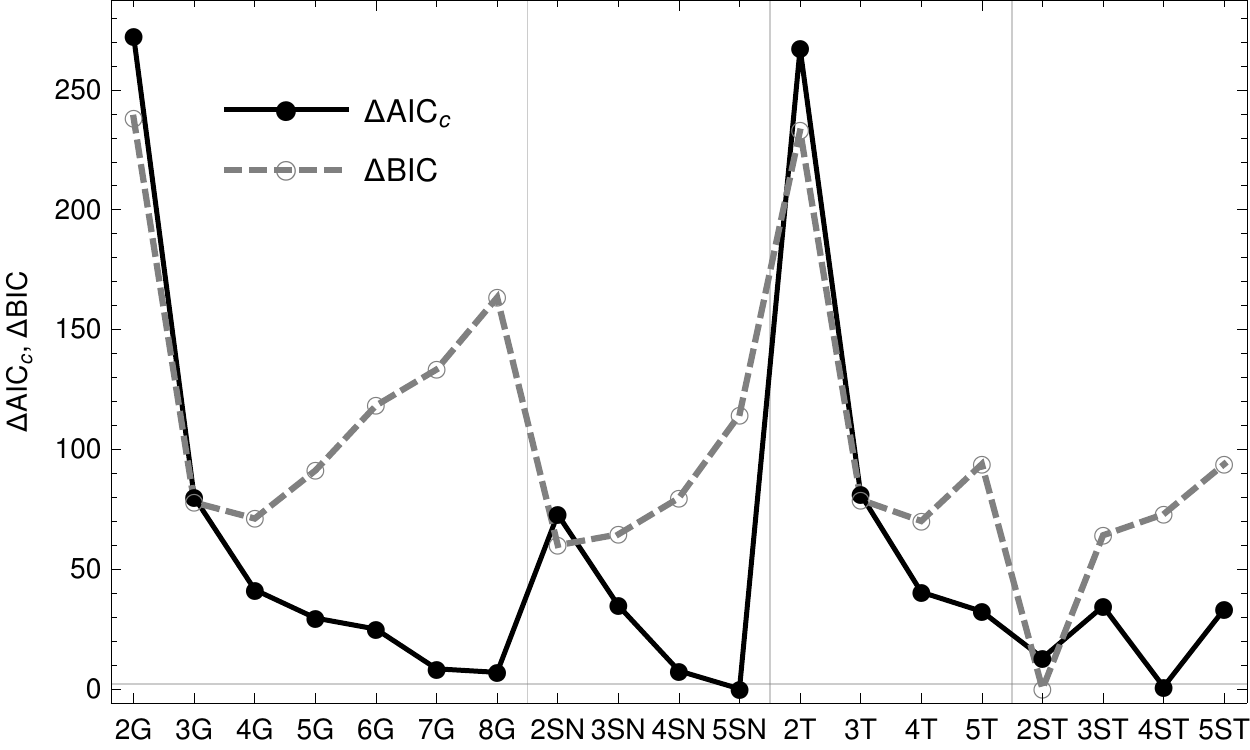}
\caption{Scatter plots and histograms (top) and information criteria scores (bottom) for $PC_1-PC_2$.}
\label{fig1}
\end{figure}

\begin{figure}
\includegraphics[width=\columnwidth]{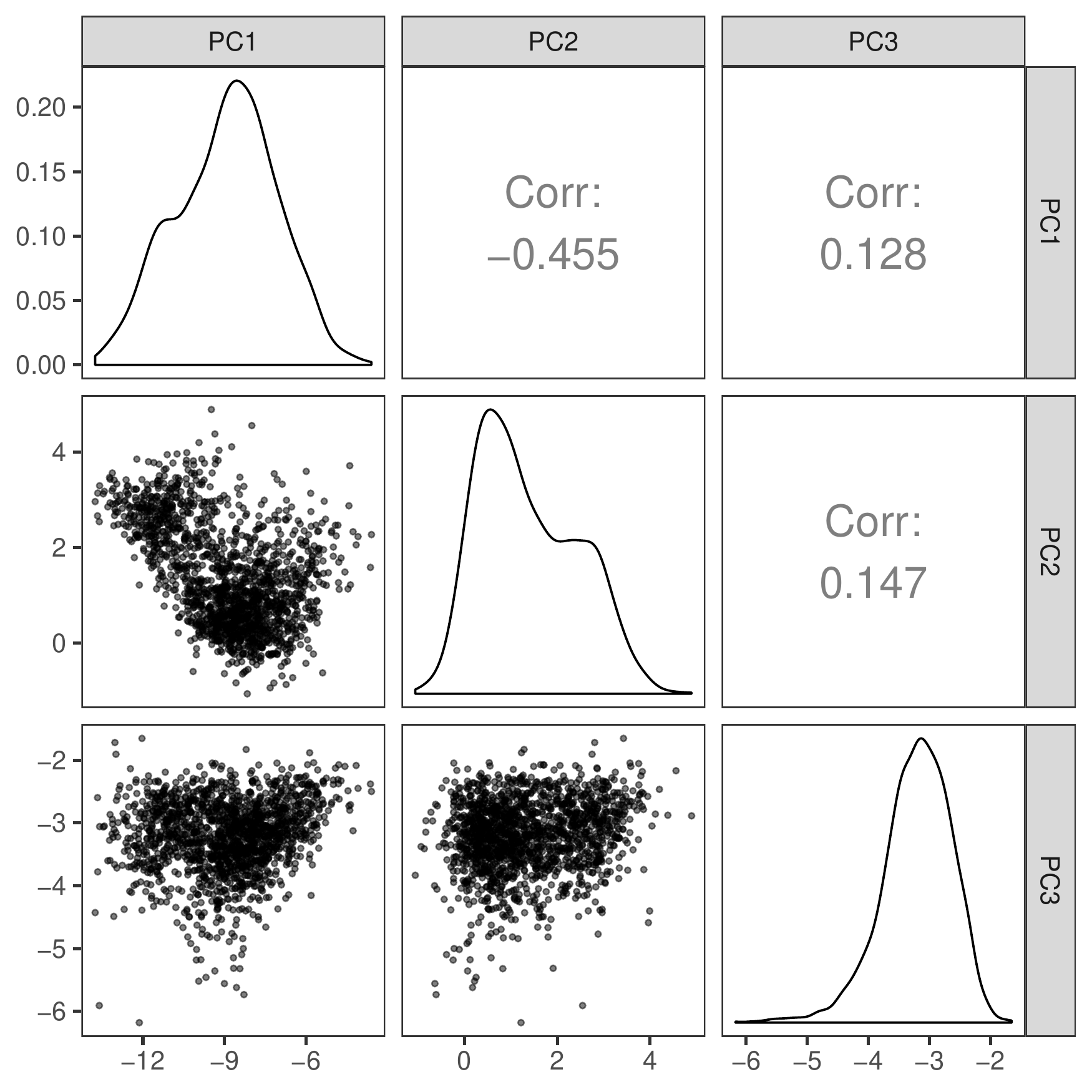}
\includegraphics[width=\columnwidth]{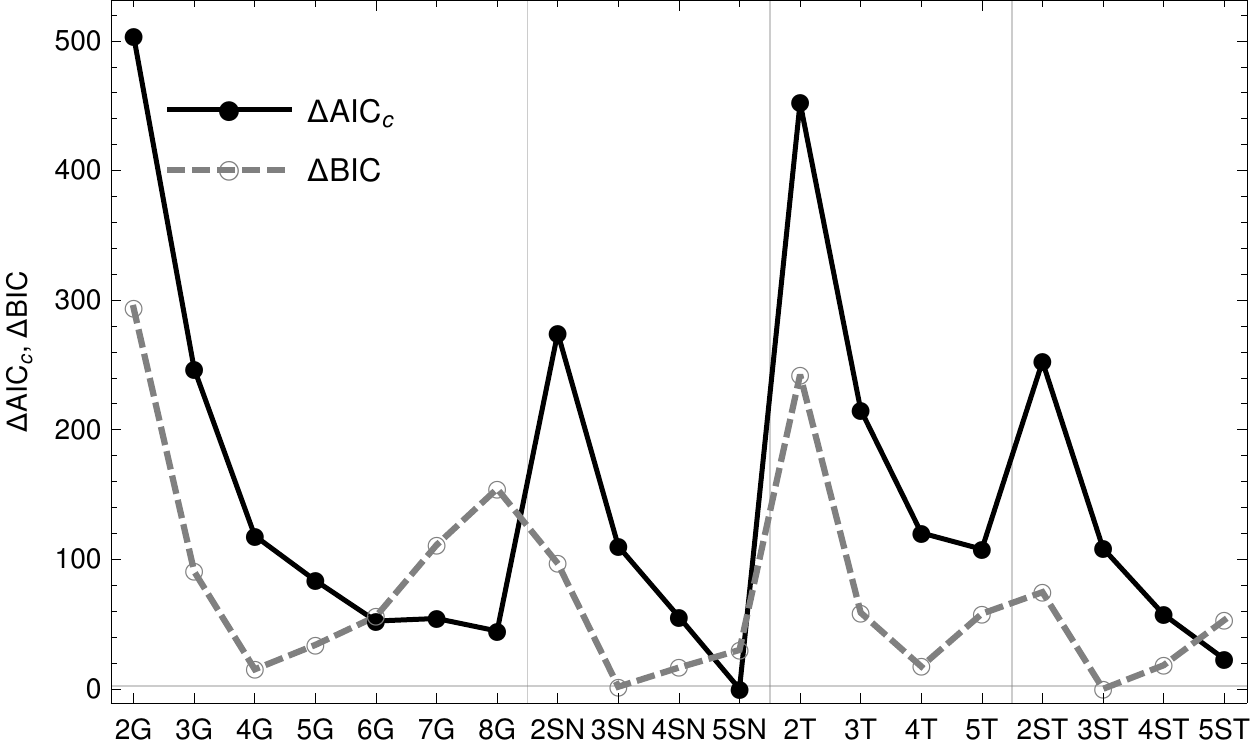}
\caption{Scatter plots and histograms (top) and information criteria scores (bottom) for $PC_1-PC_2-PC_3$.}
\label{fig2}
\end{figure}

\begin{figure}
\includegraphics[width=\columnwidth]{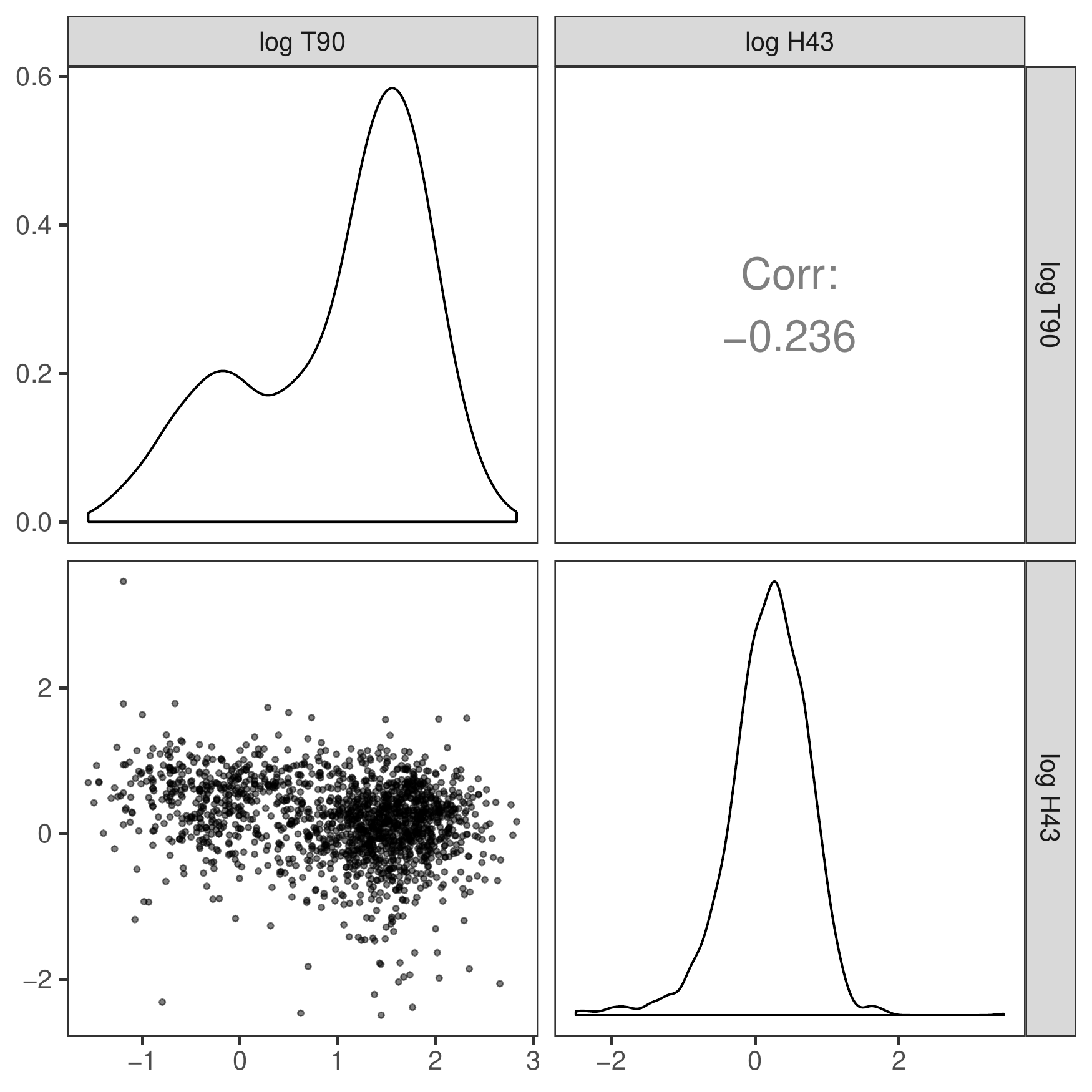}
\includegraphics[width=\columnwidth]{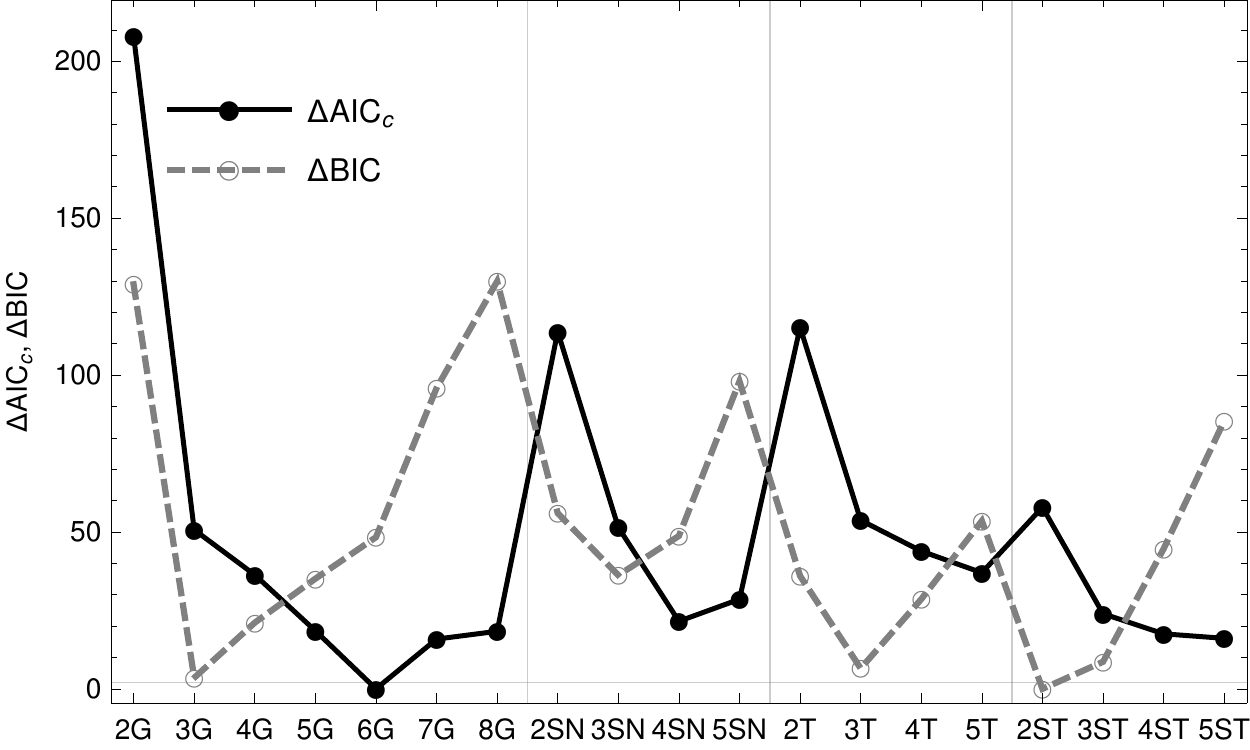}
\caption{Scatter plots and histograms (top) and information criteria scores (bottom) for $\log T_{90} - \log H_{43}$.}
\label{fig3}
\end{figure}

\begin{figure}
\includegraphics[width=\columnwidth]{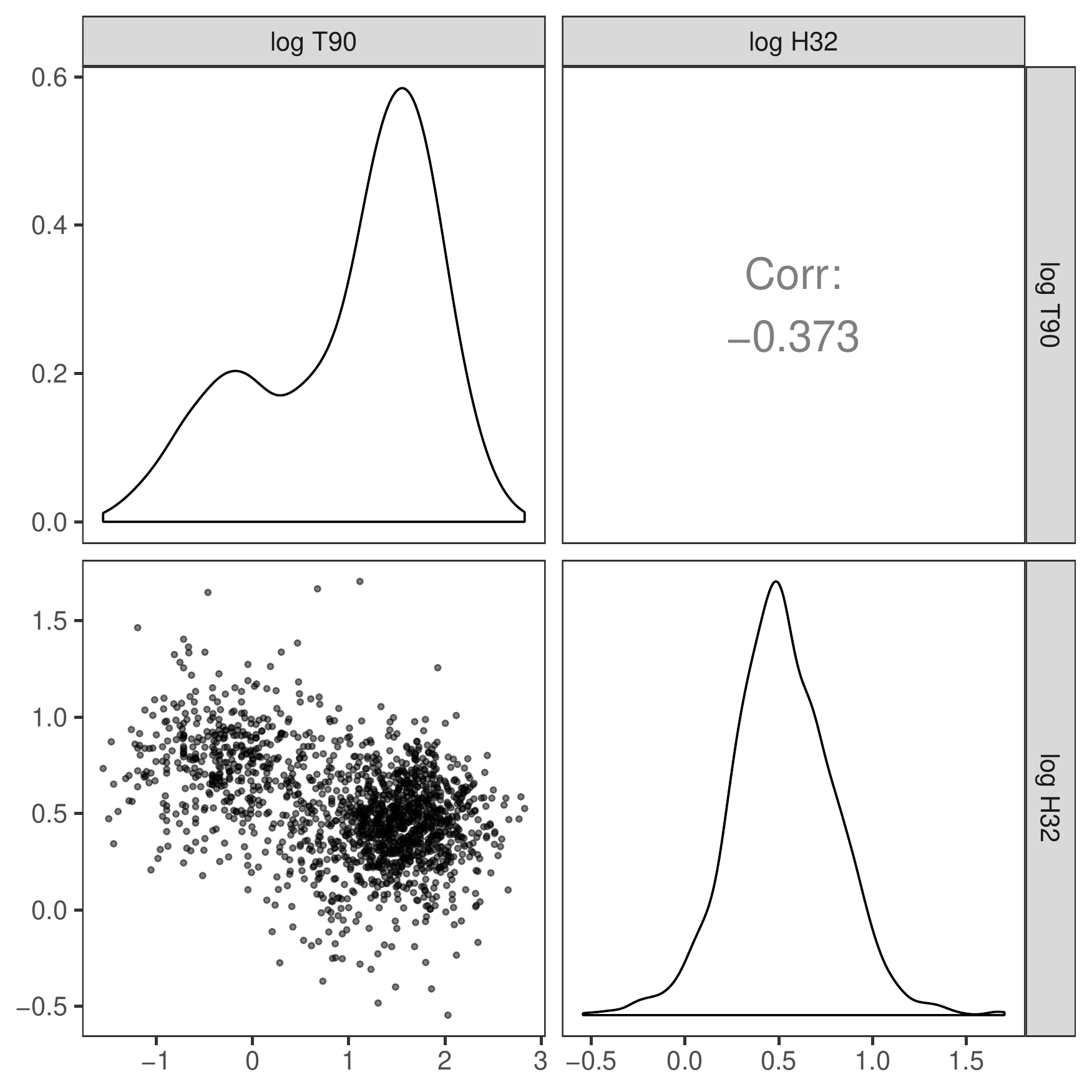}
\includegraphics[width=\columnwidth]{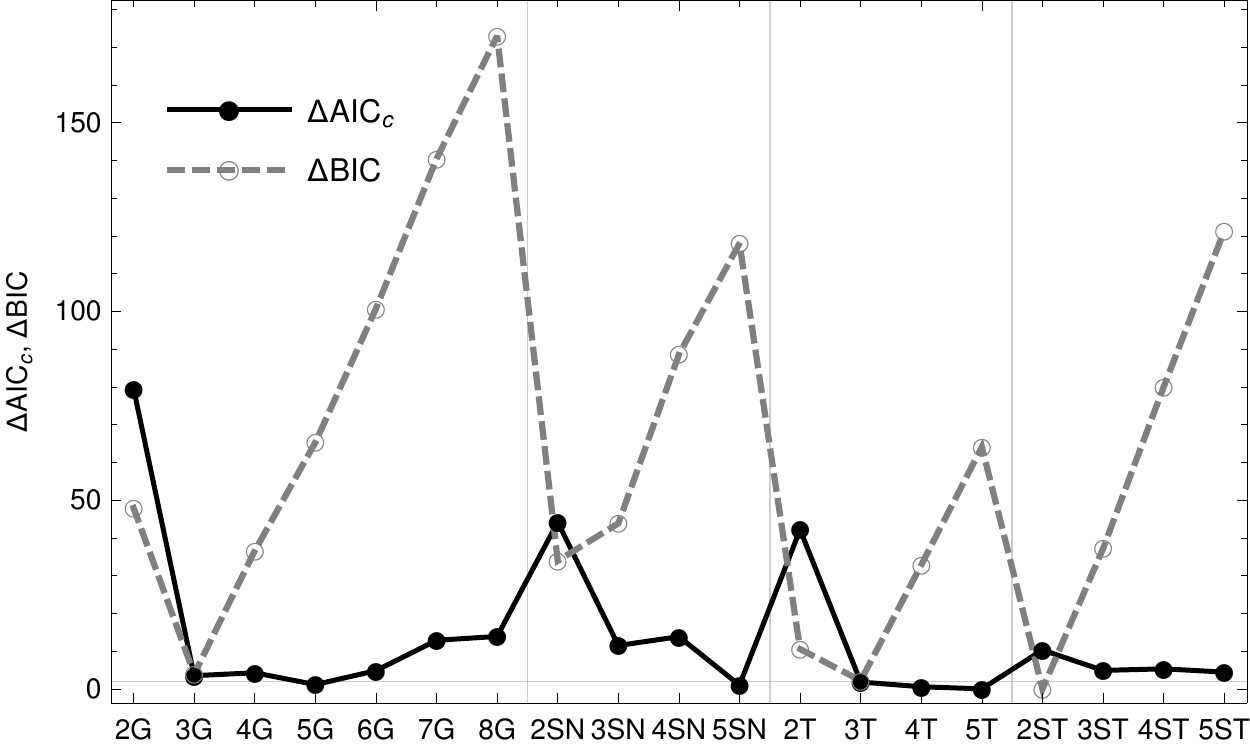}
\caption{Scatter plots and histograms (top) and information criteria scores (bottom) for $\log T_{90} - \log H_{32}$.}
\label{fig4}
\end{figure}

\begin{figure}
\includegraphics[width=\columnwidth]{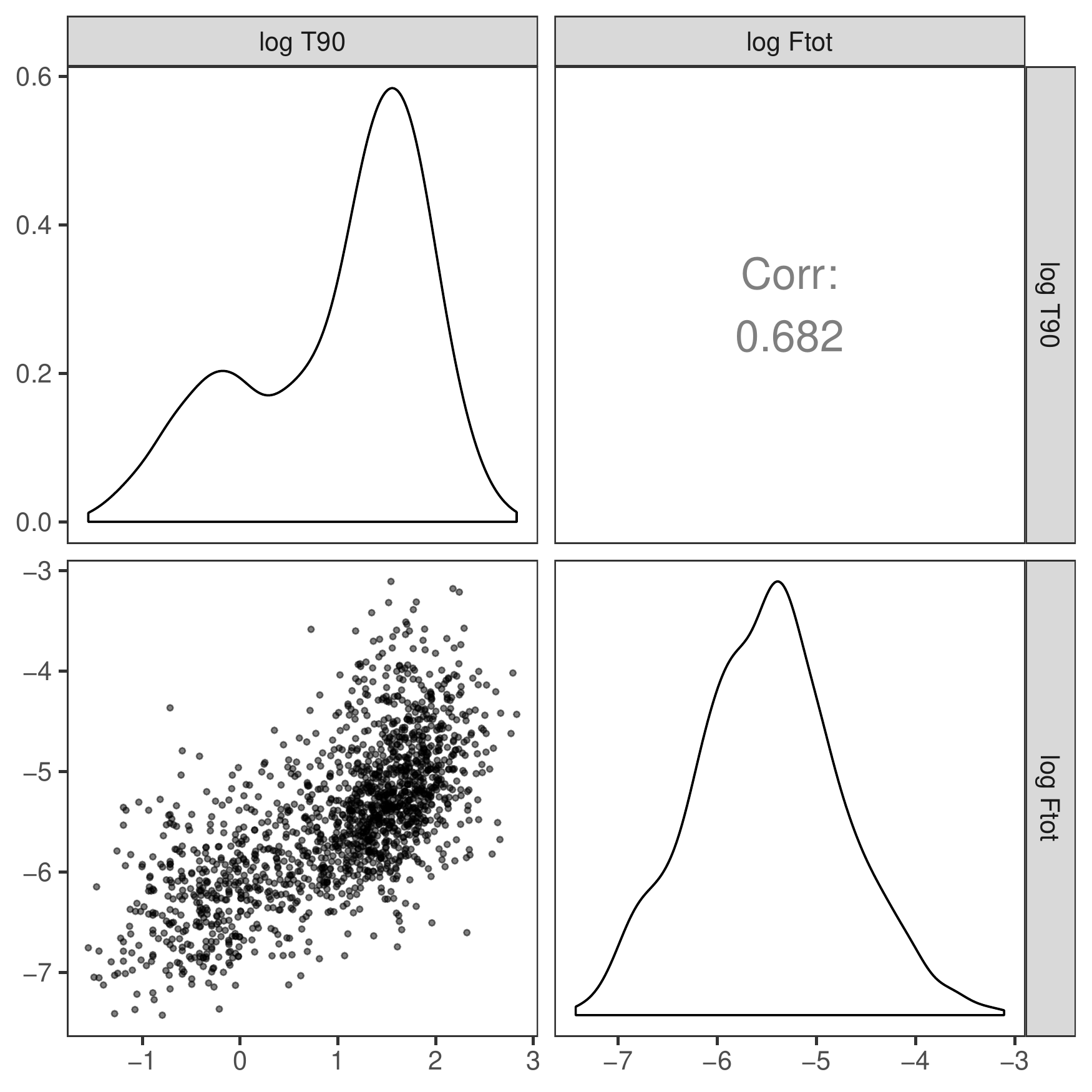}
\includegraphics[width=\columnwidth]{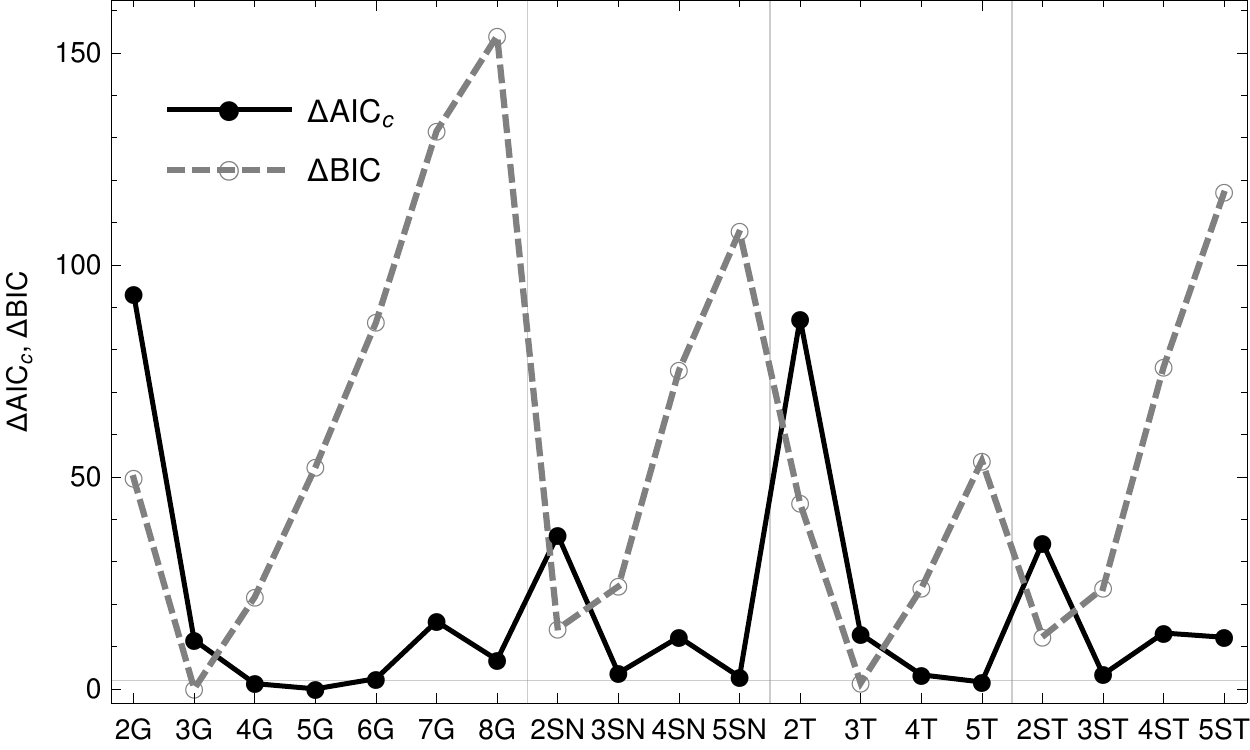}
\caption{Scatter plots and histograms (top) and information criteria scores (bottom) for $\log T_{90} - \log F_{\rm tot}$ with $F_4\neq 0$.}
\label{fig5}
\end{figure}

\begin{figure}
\includegraphics[width=\columnwidth]{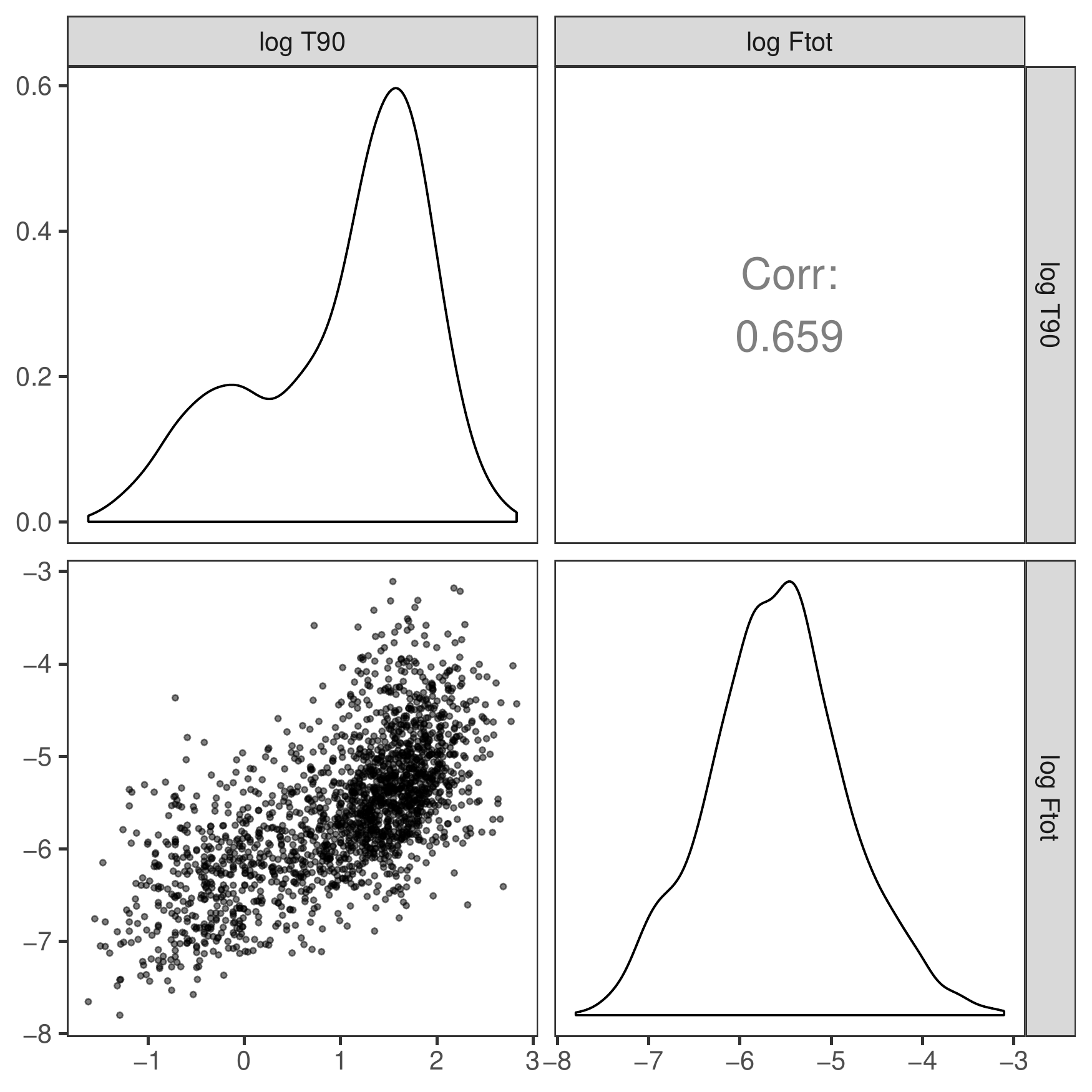}
\includegraphics[width=\columnwidth]{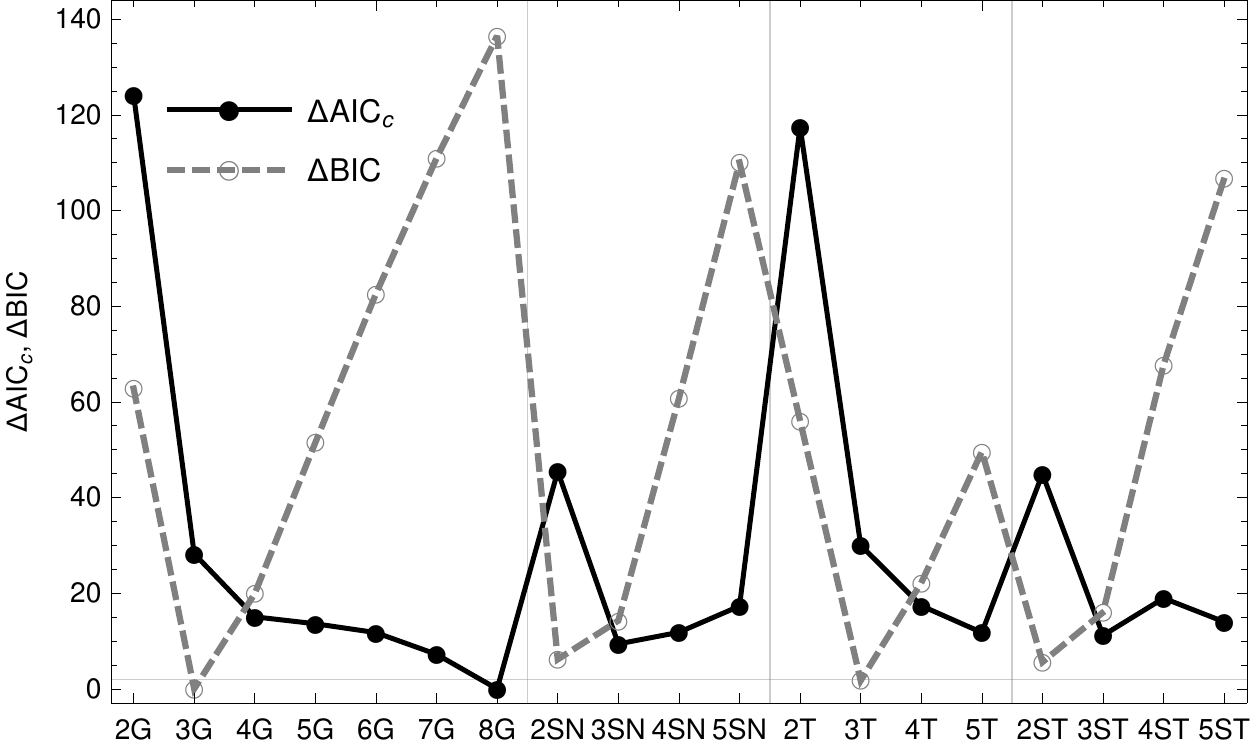}
\caption{Scatter plots and histograms (top) and information criteria scores (bottom) for $\log T_{90} - \log F_{\rm tot}$ including $F_4 = 0$.}
\label{fig6}
\end{figure}

\begin{figure}
\includegraphics[width=\columnwidth]{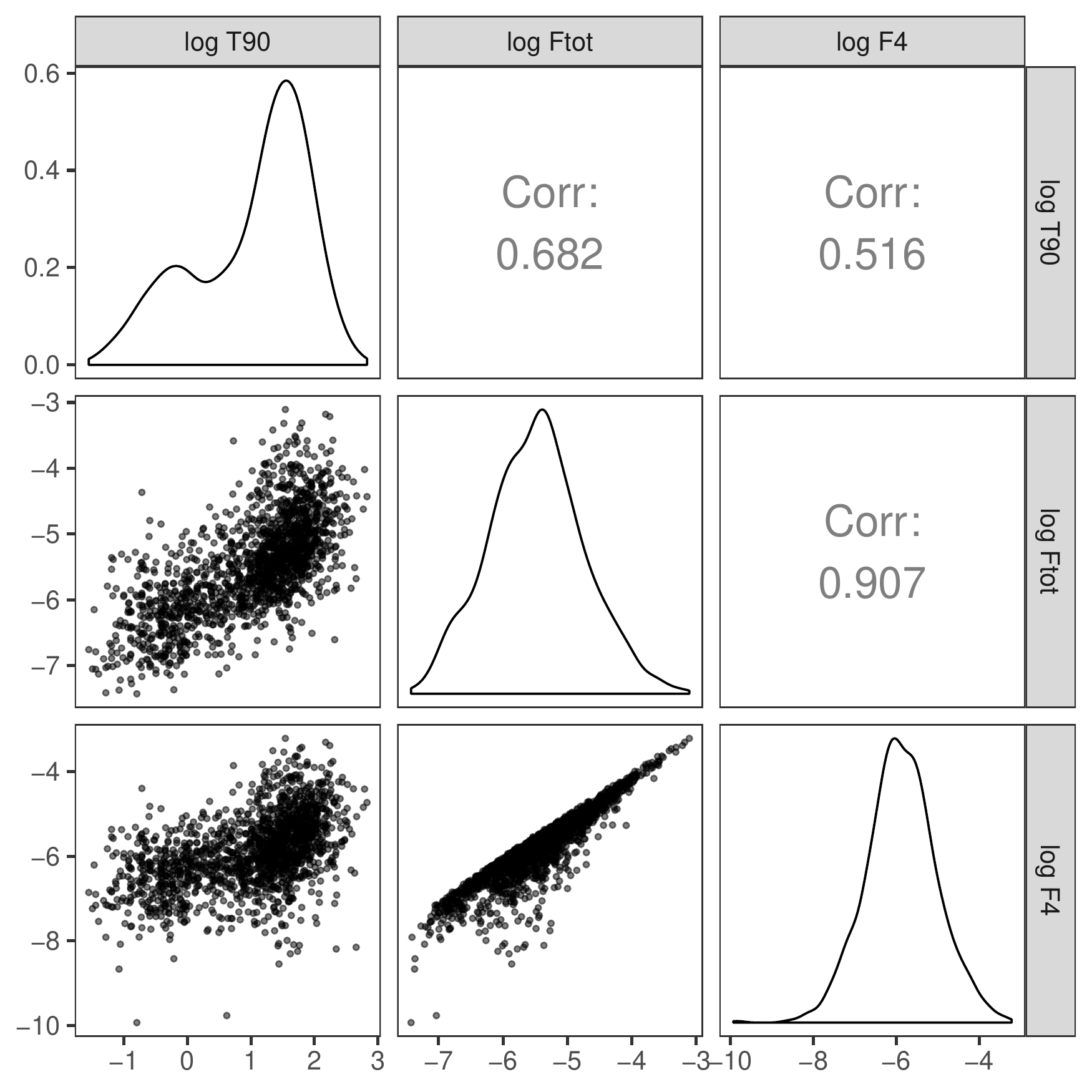}
\includegraphics[width=\columnwidth]{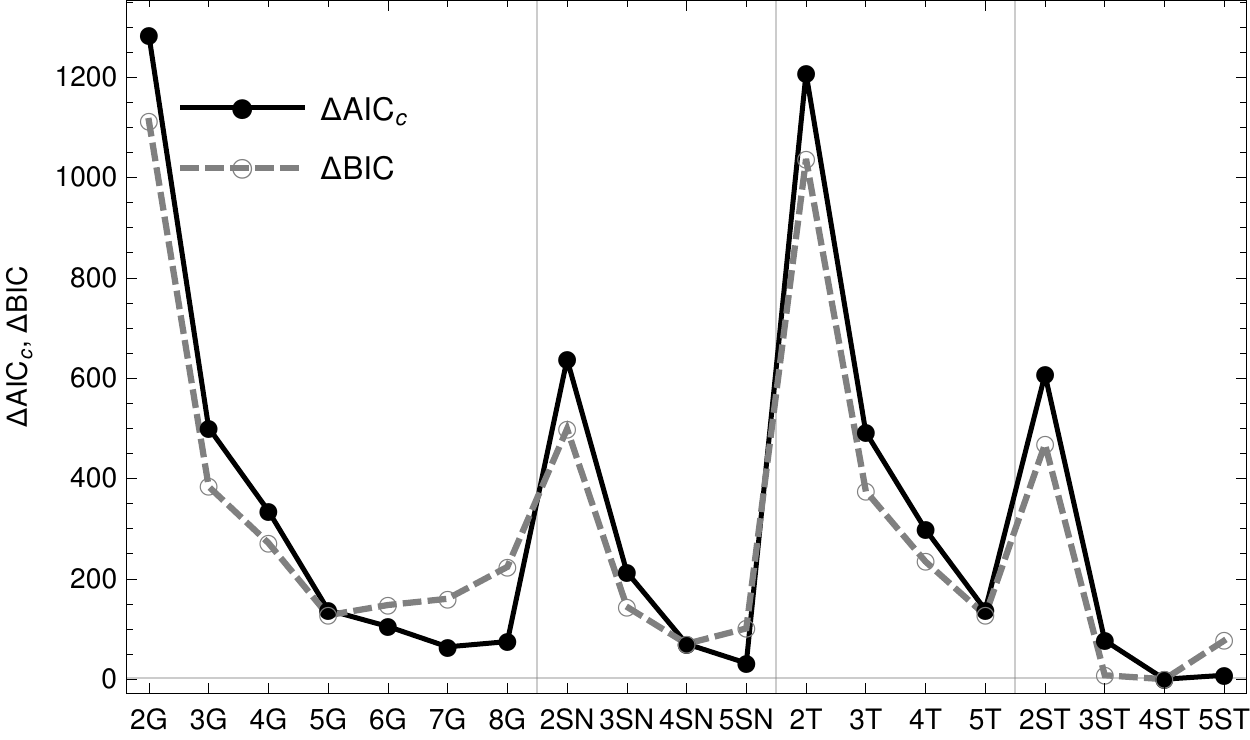}
\caption{Scatter plots and histograms (top) and information criteria scores (bottom) for $\log T_{90} - \log F_{\rm tot} - \log F_4$.}
\label{fig7}
\end{figure}

\begin{figure}
\includegraphics[width=\columnwidth]{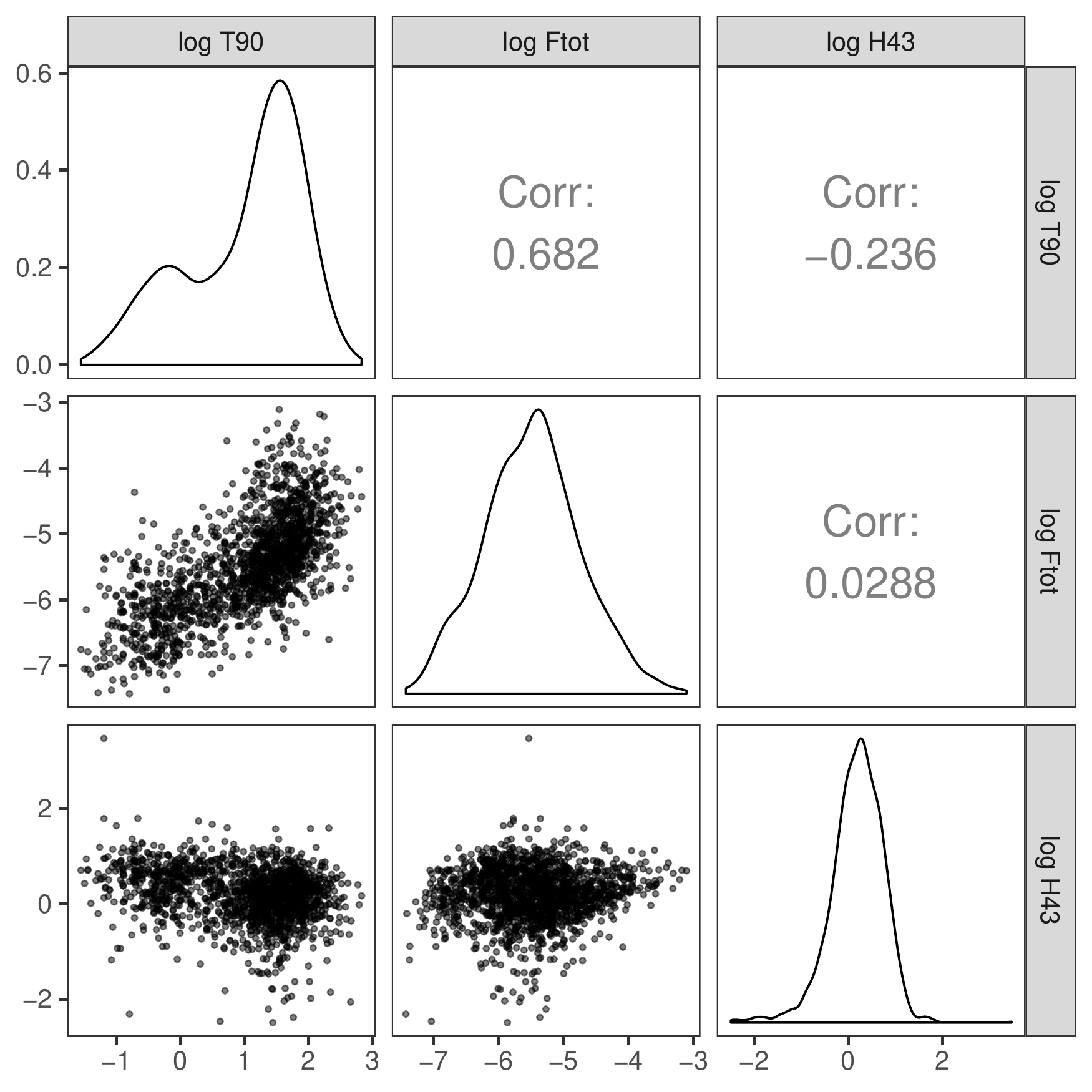}
\includegraphics[width=\columnwidth]{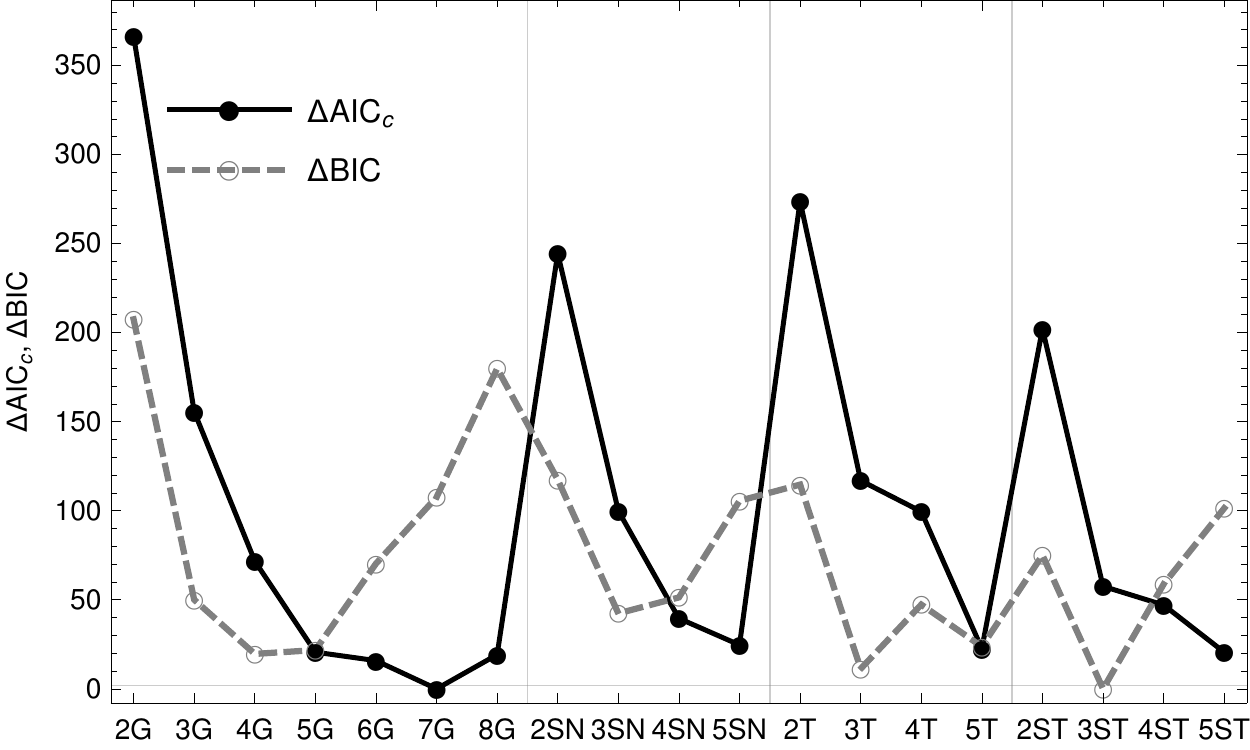}
\caption{Scatter plots and histograms (top) and information criteria scores (bottom) for $\log T_{90} - \log F_{\rm tot} - \log H_{43}$.}
\label{fig8}
\end{figure}

\begin{figure}
\includegraphics[width=\columnwidth]{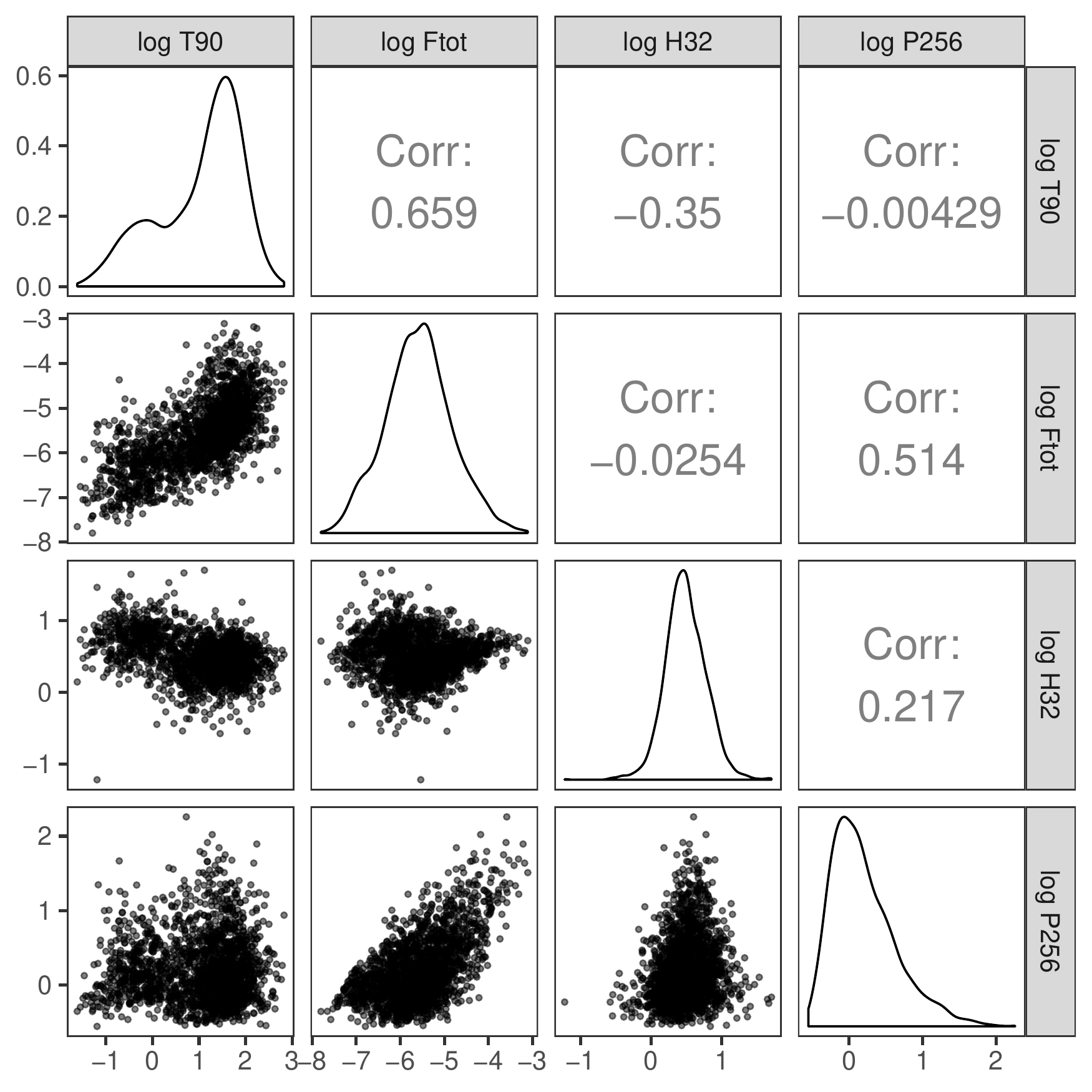}
\includegraphics[width=\columnwidth]{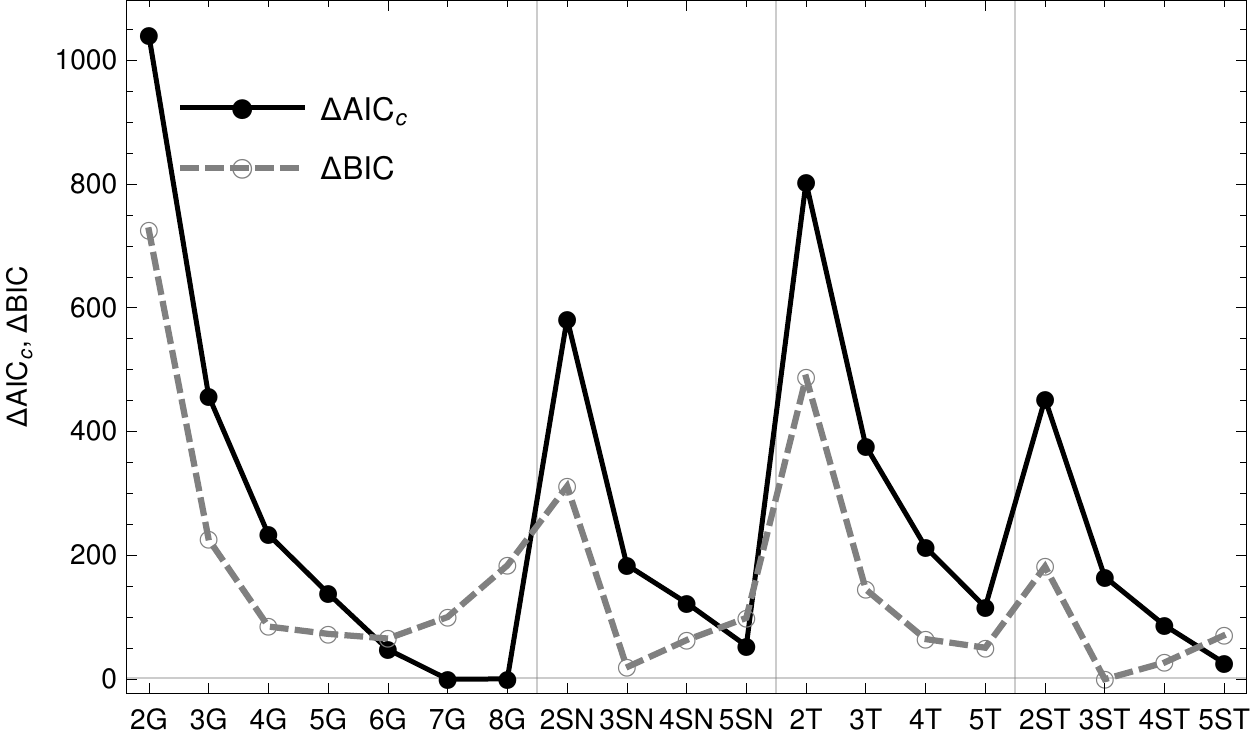}
\caption{Scatter plots and histograms (top) and information criteria scores (bottom) for $\log T_{90} - \log F_{\rm tot} - \log H_{32} - \log P_{256}$.}
\label{fig9}
\end{figure}

The samples described in Sect.~\ref{sect2.1} were fitted with the models from Sect.~\ref{sect2.2}. The number of components in the multinormal mixtures spanned 2--8 (so the number of parameters to be fitted ranges from $p=11$ to $p=47$), and 2--5 components were considered for the remaining distributions (i.e., SN, T, and ST, with $p$ reaching 40 for 5ST; see Sect.~\ref{sect2.2}). The results are displayed in Figs.~\ref{fig0}--\ref{fig9}, together with scatter plots and histograms presenting the data, and Pearson correlation coefficients. The best-fitting models (within the band $\Delta_i < 2$) are gathered in Table~\ref{tbl1}.

A few trends can be observed. First, $AIC_c$ points at quite complicated models (e.g., 5SN or 8G) compared to $BIC$. This is, however, not a surprising behavior, as $AIC_c$ tends to overfit, while $BIC$ underfits. Second, $AIC_c$ allows more complex models when the dimensionality of the data is high, e.g. for a univariate $\log T_{90}$ distribution (Fig.~\ref{fig0}) 3T is the most complex model pointed at, while the four-dimensional space of $\log T_{90} - \log F_{\rm tot} - \log H_{32} - \log P_{256}$ (Fig.~\ref{fig9}) appears to be best described by 8G. Next, only in case of $\log T_{90} - \log F_{\rm tot} - \log F_4$ (Fig.~\ref{fig7}) both $AIC_c$ and $BIC$ yield the same model, 4ST. For all other data sets there are discrepancies, with differences as big as 6G vs. 2ST for $\log T_{90} - \log H_{43}$ (Fig.~\ref{fig3}), 8G vs. 3G/3T for $\log T_{90} - \log F_{\rm tot}$ (including $F_4 = 0$, Fig.~\ref{fig6}), or 8G/7G vs. 3ST for $\log T_{90} - \log F_{\rm tot} - \log H_{32} - \log P_{256}$ (Fig.~\ref{fig9}). 

Based on the results of benchmark testing from Sect.~\ref{sect3}, the $BIC$ appears to be more reliable than $AIC_c$ in such applications. In particular, $BIC$ was 100\% accurate in detecting a 2G in the generated data, and yielded a higher accuracy than $AIC_c$ when 2SN was the true model (73\% vs. 66\%, respectively). Therefore, given such a strong tendency to overfit that was observed for the examined samples of BATSE GRBs, $AIC_c$ turns out not to be very useful, and hence $BIC$ should be used as a sole model discriminator for higher dimensional data sets. 

\begin{table*}
\caption{Best-fitting models (within $\Delta_i<2$).}
\label{tbl1}
\centering
\begin{tabular}{c c c c}
\hline
Fig. & Data & Best models ($AIC_c$) & Best models ($BIC$) \\
\hline
\ref{fig0} & $\log T_{90}$ (2037 GRBs) & 2ST, 3G, 3T & 2G, 2T \\

\ref{fig1} & $PC_1-PC_2$ (1598 GRBs) & 5SN, 4ST & 2ST \\
\ref{fig2} & $PC_1-PC_2-PC_3$ (1598 GRBs) & 5SN & 3ST, 3SN \\

\ref{fig3} & $\log T_{90} - \log H_{43}$ (1598 GRBs) & 6G & 2ST \\
\ref{fig4} & $\log T_{90} - \log H_{32}$ (1597 GRBs) & 5T, 4T, 5SN, 5G, 3T & 2ST \\

\ref{fig5} & $\log T_{90} - \log F_{\rm tot}$ (1598 GRBs) & 5G, 4G, 5T & 3G, 3T \\
\ref{fig6} & $\log T_{90} - \log F_{\rm tot}$ (1927 GRBs) & 8G & 3G, 3T \\

\ref{fig7} & $\log T_{90} - \log F_{\rm tot} - \log F_4$ (1598 GRBs) & 4ST & 4ST \\
\ref{fig8} & $\log T_{90} - \log F_{\rm tot} - \log H_{43}$ (1598 GRBs) & 7G & 3ST \\

\ref{fig9} & $\log T_{90} - \log F_{\rm tot} - \log H_{32} - \log P_{256}$ (1927 GRBs) & 7G, 8G & 3ST \\
\hline
\end{tabular}
\end{table*}

\section{Discussion and summary}
\label{sect5}

Multivariate modeling using skewed distributions was performed herein on the BATSE data set in various parameter spaces, ranging from the univariate $\log T_{90}$ case, through several two- and three-dimensional ones, e.g. spanned by the PCs, up to a four-dimensional $\log T_{90} - \log F_{\rm tot} - \log H_{32} - \log P_{256}$ instance (see Table~\ref{tbl1}). The six-dimensional case \citep{mukh,chattopadhyay17,toth19}, i.e. with added $T_{50}$ and $H_{321}$, was not taken into account due to very high correlations ($r>0.96$) between $T_{90}$ and $T_{50}$, and between $H_{32}$ and $H_{321}$. To assess the goodnes-of-fit, information criteria ($AIC_c$ and $BIC$) were employed, and to validate the reliability of the inferences, a Monte Carlo benchmark test was done: it turned out that realizations of two-component mixtures of three-dimensional skewed distributions might be incorrectly recognized to come from a (spuriously) greater number of components. Additionally, $AIC_c$, due to its liberal tendency to overfit, has a higher chance of accepting more complex models, e.g. with redundant components, than necessary. Therefore, focus should be laid on interpreting the results via the values of $BIC$, which has a higher than $AIC_c$ probability of correctly assessing the fit. With that in mind, the key results can be summed up as follows:
\begin{enumerate}
\item the distribution of $\log T_{90}$ comprises of two components (supporting previous results: \citealt{tarnopolski16a,kwong});
\item most two-dimensional instances are well described by two-component mixtures as well, except for those containing $F_{\rm tot}$, which yielded three-component symmetric fits (3G or 3T); in particular, previous results for the $\log T_{90} - \log H_{32}$ plane were confirmed \citep{tarnopolski19a};
\item most three-dimensional spaces yielded mixtures of three skewed components, except for the case of $\log T_{90} - \log F_{\rm tot} - \log F_4$, for which both $AIC_c$ and $BIC$ pointed at 4ST. This result possibly comes from a very high correlation ($r=0.907$) between $\log F_{\rm tot}$ and $\log F_4$, making this result likely flawed. After dropping $F_4$, this reduces to analyzing the case of $\log T_{90} - \log F_{\rm tot}$;
\item the four-dimensional case of $\log T_{90} - \log F_{\rm tot} - \log H_{32} - \log P_{256}$ indicates the presence of three skewed components.
\end{enumerate}

It therefore can not be undoubtfully concluded whether the BATSE sample consists of two or three groups, as the three-component fits can be spurious artifacts owing to the finiteness of the sample and, more importantly, to a particular realization of the random sample that lead to biased results. Also, spaces with higher dimensionality are more and more capacious, hence the identification of an excessive number of clusters might be more likely to happen, which appears to be the case here. Moreover, the distributions tested herein are in fact arbitrary, and the correct, underlying distribution, while inherently skewed, might be of a different shape. Henceforth, an exact derivation of the distributions of the variables of interest, starting from the univariate $\log T_{90}$ one, is desired, and expected to shed light on the issue. Also, as initially put forward by \citet{tarnopolski19a}, the redshift distribution might play a crucial role in explaining the observed skewness.

The nature of the putative intermediate GRBs is unclear. The overdensity, manifesting through emergence of a third component in the Gaussian mixture modeling, was considered previously to consist of X-ray flashes \citep{veres,grupe13}, but this cannot be a universal explanation \citep{ripa14,ripa16} due to inconsistent hardness ratios of the presumed third groups in BATSE and RHESSI (\citealt{ripa12}; see also the discussion in \citealt{tarnopolski19a}). As sGRBwEE are a plausible link between short and long GRBs \citep{kaneko15}, with their typical durations of $\sim 5-100\,{\rm s}$, they are attractive potential candidates for the intermediate group (see also \citealt{kann}).

In the end, the existence of this elusive, third class of GRBs remains debatable. As implied by benchmark testing, and the inconsistency of the outcomes obtained for the various spaces examined herein, a phenomenological approach of fitting various distributions and seeking the best model does not provide unambiguous interpretations. In particular, very different fits, with the same goodness-of-fit measures, can be found for the same data \citep{koen}. Due to the methodology of statistical inference, even finding a mixture with three (or more) components (including skewed) to be the best description of a given sample, is not decisive in asserting the number of physically valid GRB groups. Therefore it is very likely that there are still in fact only two main groups of GRBs. The excess, observed via skewness of one component or introduction of a third one, might be either a statistical artifact owing to e.g. selection effects, be a result of an inherently skewed distribution governing the prompt phase, or come from cosmological dilation stretching a Gaussian distribution into a skewed one. Even in parameter spaces of dimension greater than two, where the $BIC$ firmly implies three components, the results of the benchmark testing invoke a careful examination of all possiblities. On one hand, three components might become visible only in richer spaces; on the other, PCA suggests two dimensions should be enough, leading to a preference of two GRB classes rather than three.

\acknowledgments
Support by the Polish National Science Center (NCN) through an OPUS Grant No. 2017/25/B/ST9/01208 is acknowledged.

\software{\textsc{R} (\url{http://www.R-project.org/}), \textsc{mixsmsn} \citep[][\url{https://cran.r-project.org/web/packages/mixsmsn/index.html}]{prates}, \textsc{Mathematica}.}

\bibliography{bibliography}

\end{document}